\documentclass[a4paper,11pt]{article}
\usepackage{jheppub} % for details on the use of the package, please
\usepackage[T1]{fontenc} % if needed

\title{\boldmath Exploring physical properties of compact stars in $f(R,T)-$gravity: An embedding approach}

\author[a,b]{Ksh. Newton Singh,}
\author[c,1]{Abdelghani Errehymy, \note{Corresponding author.}}
\author[b]{Farook Rahaman}
\author[d,e]{and Mohammed Daoud}

\affiliation[a]{Department of Physics, National Defence Academy, Khadakwasla, Pune-411023, India.}
\affiliation[b]{Department of Mathematics, Jadavpur University, Kolkata-700032, India.}
\affiliation[c]{Laboratory of High Energy Physics and Condensed Matter (LPHEMaC), Department of Physics, Faculty of Sciences A\"{i}n Chock, University of Hassan II, B.P. 5366 M\^{a}arif, Casablanca 20100, Morocco.}
\affiliation[d]{Department of Physics, Faculty of Sciences, University of Ibn Tofail, B.P. 133, Kenitra 14000, Morocco.}
\affiliation[e]{Abdus Salam International Centre for Theoretical Physics, Miramare, Trieste 34151, Italy.}

\emailAdd{ntnphy@gmail.com}
\emailAdd{abdelghani.errehymy@gmail.com}
\emailAdd{rahaman@associates.iucaa.in}
\emailAdd{m$\_$daoud@hotmail.com}

\abstract{Solving field equations exactly in $f(R,T)$ gravity is one of the difficult task. To do so, many authors have adopted different methods such as assuming both the metric functions, an equation of state (EoS) and a metric function etc. However, such methods may not always lead to well-behaved solutions and thereby rejection of the solutions may happen after complete calculations. Indeed, very recent works on embedding class one methods suggested that the chances of arriving at the well behaved-solution is very high thereby inspired us to used it. In class one approach, we have to ansatz one of the metric potentials and the other can be obtain from the Karmarkar condition. In this paper, we are proposing new class one solution which is well-behaved in all physical points of view. We have analyzed the nature of the solution by tuning the $f(R,T)-$coupling parameter $\chi$ and found that the solution results into stiffer EoS for $\chi=-1$ than $\chi=1$. This is because for lesser values of $\chi$, velocity of sound is more, higher $M_{max}$ in $M-R$ curve and the EoS parameter $\omega$ is larger. The solution satisfy the causality condition, energy conditions, stable and static under radial perturbations (static stability criterion) and in equilibrium (modified TOV-equation). The resulting $M-R$ diagram from this solution is well fitted with observed values of few compact stars such as PSR J1614-2230, Vela X-1, Cen X-3 and SAX J1808.4-3658. Therefore, for different values of $\chi$, we have predicted the corresponding radii and their respective moment of inertia from the $M-I$ curve.}

\begin{document} 
\maketitle
\flushbottom

\section{Introduction}\label{sec1}
One of the greatest challenges in modern cosmology is the late-time comportment of our Universe. We use several sets of high-precision observational data gathered from various cosmic sources such as Cosmic Microwave Background (CMB)~\citep{Bennett2003,Spergel2003,Spergel2007}, SuperNova type Ia (SNe Ia)~\citep{Perlmutter1997,Perlmutter1998,Perlmutter1999,Riess2004,Riess2007}, Large Scale Structure (LSS)~\citep{Hawkins2003,Tegmark2004,Cole2005}, Weak Lensing (WL)~\citep{Jain2003} and Baryon Acoustic Oscillations (BAO)~\citep{Eisenstein2005}, as standard candles which discovered that our Universe is undergoing an accelerated expansion. The fascinating part is that the Universe expansion is believed to be made from an obscure energy named dark energy, which is around two-thirds of the complete energy budget of the Universe. On account of the puzzling nature of the dark sector (Dark Energy (DE) and Dark Matter (DM); DE yields a late-time speeding up of the cosmological foundation while DM carries on as an undetectable residue matter supporting the procedure of gravitational clustering.), and the way that their existence is construed only through their gravitational impacts, it is entirely to verify whether there is a need to contemplate these elements; Specifically, regardless of whether there is any deflection from ordinary General Relativity (GR) on enormous scales. By utilizing the Einstein’s Field Equations (EFEs), there exists an accelerated expansion portrayed by a positive constant, which is extremely little, in the edge work of GR, named the $\Lambda -$CDM model~\citep{Bahcall1999}. In the present situation, this little positive constant is related with dark energy in void space, which is utilized to clarify the ongoing case of coeval accelerating expansion of the Universe against the alluring impacts of gravity. In widespread, there are two main methodologies that could supply a surroundings for a theoretical clarification of the accelerated expansion of the Universe. The first methodology be made up of altering the matter substance of the Universe by introducing a DE area, beginning either with a phantom field, a standard scalar field, or with the mixture of the two fields in a unified model and then progressing towards more complicated scenarios; see ~\citep{Copeland2006,Cai2010} and references therein for more subtleties and audits. The second methodology consists in modifying the gravitational area itself (see e.g.~\citep{Capozziello2011a,De Felice2010,Nojiri2011,Lobo2008}), which can likewise be well-respected as one of the great candidate for clarifying the accelerated expansion of the Universe. Prompted by this basic hypothesis that at great astrophysical and cosmological scales, usual GR may not depict effectively the dynamical evolution of the Universe. To manage this problem, several endeavors have been made, among which gravity theories broadening GR have aroused much enthusiasm over the previous decades. With regards to modified gravity theories, a geometric depiction for DM can be given and the accelerated expansion can be accomplished at late-times, in this way, the cosmological constant issue might be settled (for an exhaustive dynamical system investigation of some cosmological models with regards to alternative gravitational theories see ~\citep{Leon2009,Leon2011,Xu2012,Xu2013,Xu2015a,Xu2015b,Leon2013,Fadragas2014,Kofinas2014,Skugoreva2015,Carloni2016}). In the midst of the various models of DE, the altered gravity models are very fascinating as they integrate a few movements of quantum and general gravity theories. Different techniques have been suggested up until now, so as to modify the gravitational action (see~\citep{Clifton2012} for more details), offering rise to various classes of alternative gravitational theories. In this regard, some leading models incorporate $f(R)$ gravity~\citep{Capozziello2008,Capozziello2009,Nojiri2009,De  Felice2010,Maartens2010,Capozziello2010,Capozziello2011b,Nojiri2011,Capozziello2012,Astashenok2013,Capozziello2015,Astashenok2015b,Jovanovic2016,Capozziello2016b,Santos2017,Astashenok2017,Chervon2018,Capozziello2018a,Capozziello2018b,Odintsov2019a,Odintsov2019b,Capozziello2019}, $f(T)$ gravity~\citep{Bohme2011,Wang2011,Daouda2011,Sharif2013,Capozziello2013}, $f(\mathcal{T})$ gravity~\citep{Bengochea2009,Linder2010}, $f(\mathcal{G})$ gravity~\citep{Bamba2010a,Bamba2010b,Rodrigues2014}, $f(R, T)$ gravity~\citep{Yousaf2016,Yousaf2016b,Correa2016,Moraes2016a,Moraes2016b,Das2016,Deb2018,Maurya2019,Sahoo2019,Shabani2018,Barrientos2018,HansrajB2018,Singh2018,Baou2018,Hansraj2018,Yousaf2018}, $f(\mathcal{T}, T)$ gravity~\citep{Pace2017,Saez-Gomez2016,Momeni2014,Junior2015,Salako2015,Nassur2015}, $f(R, \mathcal{G})$ gravity~\citep{Nojiri2005}, where $R$, $T$, $\mathcal{T}$ and $\mathcal{G}$ are Ricci’s scalar, trace of stress-energy tensor, torsion scalar and Gauss-Bonnet scalar respectively. In $f(R)$ gravity theory, to a greater extent general expression of the Ricci scalar $R=g^{\mu \nu}R_{\mu \nu}$ is utilized instead of $R$ though $f(T)$ gravity is a general type of teleparallel gravity. Spurred by the achievement of cosmological constant as a straightforward and great candidate of DE and DM, a few matter field is additionally combined with the expression of the Ricci scalar $R$ in the action geometry sector in a few alternatives theories of gravitation ($f(R, \mathcal{L}_m)$ theory, where $\mathcal{L}_m$ is the matter Lagrangian density).

In the context of interest of including certain components of matter into the geometry of action, the $f(R, T)$ theory was suggested by \cite{har11} which, generally, has been a fascinating framework for studying acceleration models. The $f(R, T)$ gravity theory generalizes $f(R)$ gravity theories by introducing the trace of stress-energy tensor added to the Ricci scalar. The validation for the reliance on $T$ originates from enlistments emerging from a few exotic fluid or quantum impacts. Actually, this enlistment perspective encompasses or connects to the recommendations mentioned,  for instance, geometrical curvature prompting matter, a geometrical portrayal of physical powers and a geometrical source for the matter substance of the Universe. In Ref. \citep{har11}, the field equations of a few specific models are introduced, and especially, scalar field models $f(R, T^{\Phi})$ are examined in detail with a concise account of their cosmological ramifications. Likewise, the motion equation of the test particle and the Newtonian boundary of this equation are additionally studied in Ref.~\citep{har11}. Until nowadays, the problems which have been explored alongside this alternative theory are the thermodynamics~\citep{Houndjo2012a,Sharif2012,Jamil2012a}, the energy conditions~\citep{Alvarenga2013a}, anisotropic cosmology~\citep{Sahoo2017,Fayaz2014}, the cosmology where the portrayal uses a helper scalar field~\citep{Houndjo2012b}, reconstruction of some cosmological models~\citep{Jamil2012b}, the wormhole solution~\citep{Moraes2017,Sahoo2018}, the scalar perturbations ~\citep{Alvarenga2013b} and some other relevant aspects~\citep{Saha2015,Momeni2015,Noureen2015,Yousaf2016}. Additionally, a more generalization of this theory has been suggested lately in Refs.~\citep{Odintsov2013,Haghani2013}.

Consequently, it is not reasonable to affirm or to refute such theories dependent on the outcomes of cosmology and contrast them with the observational datum, for instance, the challenge of the viability of $f(R, T)$ as an alternative modification of gravity that discussed in~\citep{Velten017}. In any case, to set up an agreeable theory of gravitation, it is essential to consider at the astrophysical level, for instance by utilizing the relativistic stellar structures. A few contentions for these modified theories originate from the presumption that relativistic stellar structures in the powerful gravitational sector could distinguish usual gravity from its generalizations. In the scenario of $f(R,T)$ gravity, an enormous number of contributions on the evolution of compact stellar structures are accessible in various literature. In this context, hydrostatic equilibrium structure of strange stars and neutron stars have been investigated~\citep{Moraes2016b}. The configuration of compact stellar structures in $f(R,T)$ gravity was explored latterly in Refs.~\citep{Zubair2016,Das2016,Deb2018,Deb2019,Maurya2019,Maurya2019b}, though gravastars (GRAvitational VAcuum STARS) resolution has been gotten in~\citep{Das2017}.

To understand the inside geometry and evolutionary phases of relativistic stellar structures, the distribution of anisotropic fluids acts an important role. As the compact stellar systems have ultra-dense cores and their density surpass the nuclear density, consequently, pressure ought to be anisotropic in the inside of compact stellar systems~\citep{Herr92}. In anisotropic relativistic astrophysical systems, it is seen that pressure is apportioned into radial and tangential components. In this specific circumstance, numerous researchers explored attributes of compact and dense stellar systems involving anisotropic fluid structure.~\citep{ruder72} first offer the anisotropy concept for static spherically symmetric structures and subsequently, numerous astrophysicists have included this parameter for the compact stellar structures modeling. As a compact stellar structure is shaped with very dense matter, a very-high magnetic domain is related with it due to the magnetic flux preservation. The enormous magnetic domain may produce pressure anisotropy interior the stellar structure~\citep{Weber99}. The stage progress at higher densities may likewise prompt anisotropy~\citep{Sokolov80, Sawyer72}. The anisotropy aspect furnishes us with one more factor to be incorporated in EFEs and prompting progressively realistic models of the spherically symmetric compact stellar structures. If the anisotropy parameter is positive an outward repulsive force will be applied on the stellar structure in this way, making it increasingly compact and steady. The primary explanation behind emerging anisotropy in a $f(R, T)$ gravity theory model could be the anisotropic type of the two fluid system without interaction. Also, it ought to be emphasized that from a quantum point of view, the $T-$dependent Lagrangian might be identified with the formation of particles which normally portray the presence of bulk viscosity and other flaws in the alluded fluid. Consequently, we suggest the models where transverse pressure surpasses the radial pressure. To provide the accurate solutions of EFEs, two dissimilar methodologies are frequently follow: it is possible that we learn the space-time metric elements first, afterward establish the matter profile, or we first portray the material features in terms of certain state equations i.e. the relationship in the form $p = p(\rho)$ and subsequently investigate the metric potentials. Whereas searching a well-defined solution the constants arising in the solution perform an important role. The slight difference in the values of the constants can distance the stellar structure from its position of equilibrium. Subsequently, several authors have been studied compact stellar structure models employing the Karmarkar condition. At the moment when the Karmarkar condition~\citep{kar48} is exploited on the gravitational components $g_{rr}$ and $g_{tt}$, the issue of establishing the spatio-temporal elements obtain make simple to a great area and the 4-dimensional Riemannian spatio-temporal variety can be described graphically in the 5-dimensional pseudo-Euclidean spatio-temporal variety without any change in its intrinsic characteristics. The solutions fulfilling Karmarkar conditions alongside the condition suggested by~\citep{pan81} are well-known as embedding class one solutions. It is intriguing to observe that the internal solution of~\citep{Sch16} is the only structure of bounded neutral matter with a disappearing anisotropy parameter fulfilling the Karmarkar condition. For a more in-depth survey, one may seek advice from alluded literature~\citep{Barnes74,Akbar017,Kuhfittig018} where authors have clearly involved and examined the impacts of the procedure of embedding of 4-dimensional Riemannian spatio-temporal variety into the 5-dimensional pseudo-Euclidean spatio-temporal variety in the scenario of GR and alternative gravity.

In this paper, we study anisotropic spherically symmetric solutions in the domain of alternative gravity theories, especially, $f(R, T)$ theory of gravity. In this respect, we have considered that the matter Lagrangian density $\mathcal{L}_m$ (defined as $\mathcal{L}_m=-\mathcal{P}=(p_r+2p_t)/3$ i.e the isotropic pressure) can be asserted as the linear function of the Ricci scalar $R$ and the trace of the energy-momentum tensor $T$, i.e. $f(R, T)=R + 2\chi T$, where $\chi$ is a dimensionless coupling constant, in order to depict the global set of modified EFEs for the anisotropic matter distribution. We also consider the embedding class I procedure, by means for embedding within 4-dimensional space-time into a 5-dimensional flat Euclidean space to obtain a complete space-time representation interior the relativistic stellar system. Moreover, for investigating physical availability of the acquired solutions, we have analyzed four different compact stars namely PSR J1614-2230, Vela X-1, Cen X-3 and SAX J1808.4-3658 linked physical parameters analytically and graphically. Accordingly, the familiar Darmois-Israel~\citep{Darmois1927,Israel1966} coordinating conditions can be used to calculate all the physical and constant ingredients of the stellar system.

The paper is organized as follows: Beginning with a brief introduction in Sec.~\ref{sec1}, we make a review of the concept $f(R,T)-$ gravity theory in Sec.~\ref{sec2}, next in Sec.~\ref{sec3}, the basic EFEs for anisotropic matter distributions in $f(R, T)-$gravity is described. In Sec.~\ref{sec4} we will show the Karmarkar condition well-known as embedding class one solution, then in Sec.~\ref{sec5} the complete stellar system under embedding class one technique in the arena of $f(R, T)-$gravity is gained and its thermodynamic description is given. In Sec.~\ref{sec6} we will analyze the new solutions through various physical tests such as hydrostatic equilibrium, causality condition, stability factor,  adiabatic index and stability, static stability criterion and energy conditions. In Sec.~\ref{sec7} we coordinate the acquired stellar system with the outside spacetime given by Schwarzschild metric, so as to get the constant parameters. Furthermore, the stiffness of EoS, $M-R$ and $I-M$ diagram are discussed in Sec.~\ref{sec8}. Finally, we conclude our investigation with a short discussion of the result in Sec.~\ref{sec9}.

\section{Concepts of $f(R, T)-$gravity}\label{sec2}

In the Einstein-Hilbert action, if one replace the Ricci scalar $R$ by a function of $R$ and the trace of stress-energy tensor $T$ we arrived at the modified action in $f(R,T)-$gravity as
\begin{eqnarray}\label{eq1}
S &=& \frac{1}{16\pi}\int f(R,T)\sqrt{-g}~ d^{4}x+\int \mathcal{L}_{m}\sqrt{-g}~d^{4}x
\end{eqnarray}
where det$(g_{\mu \nu})=g$. The source term of the matter Lagrangian density $\mathcal{L}_m$ defines a stress tensor as
\begin{eqnarray}
T_{\mu\nu}=-\frac{2}{\sqrt{-g}}\,\frac{\delta \big(\sqrt{-g}\,\mathcal{L}_m \big)}{\delta g^{\mu\nu}}. \label{eq2}
\end{eqnarray}
Following \cite{har11} approach, the Eqs. \eqref{eq2} reduce to
\begin{eqnarray}
T_{\mu\nu}=g_{\mu\nu} \mathcal{L}_m-2\frac{\partial \mathcal{L}_m}{\partial g^{\mu\nu}}. \label{eq3}
\end{eqnarray}

Variation of the action w.r.t $g_{\mu \nu}$ implies the field equations
\begin{eqnarray}\label{eq4}
&&\left( R_{\mu\nu}- \nabla_{\mu} \nabla_{\nu} \right)f_R (R,T) +g_{\mu\nu} \Box f_R (R,T) - \frac{1}{2} f(R,T)g_{\mu\nu}  \nonumber \\
&& \hspace{6 cm}= 8\pi ~T_{\mu\nu} - f_T (R,T)~ \Big( T_{\mu\nu}+\Theta_{\mu \nu} \Big),
\end{eqnarray}
provided $f_R (R,T)={\partial f(R,T)}/{\partial R}$ and $f_T (R,T)={\partial f(R,T)}/{\partial T}$. The $\nabla_\mu$ denotes covariant derivative while thye box operator $\Box$ is defined by
\begin{eqnarray}
\Box \equiv {1 \over \sqrt{-g}} {\partial \over \partial x^\mu}\left(\sqrt{-g}~g^{\mu\nu} {\partial \over \partial x^\nu} \right) ~~ \textrm{with} ~~ \Theta_{\mu\nu}=g^{\alpha\beta} {\delta T_{\alpha\beta} \over \delta g^{\mu\nu}}. \nonumber
\end{eqnarray}
The conservation equation \citep{bar14} yields
\begin{eqnarray}\label{eq5}
\nabla^{\mu}T_{\mu\nu}&=&\frac{f_T(R, T)}{8\pi -f_T(R,T)}\bigg[(T_{\mu\nu}+\Theta_{\mu\nu})\nabla^{\mu}\ln f_T(R,T)   +\nabla^{\mu}\Theta_{\mu\nu}-\frac{1}{2}g_{\mu\nu}\nabla^{\mu}T\bigg].
\end{eqnarray}
Therefore, $\nabla^{\mu}T_{\mu\nu} \neq 0$ implies the conservation equation no longer holds in $f(R,T)$ theory. By using Eq. (\ref{eq3}), the tensor $\Theta_{\mu\nu}$ is found to be
\begin{equation}\label{eq6}
\Theta_{\mu\nu}= - 2 T_{\mu\nu} +g_{\mu\nu}\mathcal{L}_m - 2g^{\alpha\beta}\,\frac{\partial^2 \mathcal{L}_m}{\partial g^{\mu\nu}\,\partial g^{\alpha\beta}}.~~~
\end{equation}

To complete the field equations, we assumed an anisotropic fluid source 
\begin{equation}\label{eq8}
T_{\mu\nu}=(\rho+p_r) u_\mu u_\nu-p_t g_{\mu\nu}+(p_r-p_t)g_{\mu \nu},
\end{equation}
with ${u_{\nu}}$ is the four velocity, satisfying $u_{\mu}u^{\mu}= -1$ and $u_{\nu}\nabla^{\mu}u_{\mu}=0$, $\rho$ is the matter density, $p_r$ and $p_t$ are the radial and transverse pressures.  If we defined the isotropic pressure as $-\mathcal{P}=\mathcal{L}_m=(p_r+2p_t)/3$ \citep{har11}, then (\ref{eq6}) reduces to
\begin{eqnarray} \label{eq9}
\Theta_{\mu\nu}=-2T_{\mu\nu}-\mathcal{P}g_{\mu \nu}.
\end{eqnarray}

Further, the functional $f(R,T)$ is chosen to be $f(R,T)= R+2\chi T$ \citep{har11}, where $\chi$ the coupling constant. Now the field equations (\ref{eq5}) takes the form 
\begin{eqnarray}\label{eq10}
G_{\mu\nu}= 8\pi ~T_{\mu\nu}+\chi T~ g_{\mu\nu}+2\chi(T_{\mu\nu}+\mathcal{P}g_{\mu\nu}).
\end{eqnarray}
For $\chi=0$, one can recover the general relativistic field equations. The linear expression in $f(R,T)$ solved many cosmological and astrophysical related problems. By substituting $f(R,T)=R+2\chi T$ and (\ref{eq9}) in Eq. (\ref{eq5}), we obtain
\begin{eqnarray}\label{eq11}
\nabla^{\mu}T_{\mu\nu}=-\frac{\chi}{2(4\pi+\chi)}\bigg[g_{\mu\nu}\nabla^{\mu}T+2\,\nabla^{\mu} \big(\mathcal{P} g_{\mu\nu} \big)\bigg].
\end{eqnarray}
Therefore, the conservation equation in Einstein's gravity can be recovered for $\chi=0$. 
\section{Field equations in $f(R,T)-$gravity}\label{sec3}

To determine the field equations we assume an interior spacetime of the form
\begin{equation}
ds_-^2 = e^\nu dt^2 - e^\lambda dr^2 - r^2 (d\theta^2 + \sin ^2 \theta ~d\phi^2). \label{met}
\end{equation} 

For the spacetime given in (\ref{met}), the field equation (\ref{eq10}) becomes
\begin{eqnarray}
8\pi \rho_{eff} &=& e^{-\lambda} \left({\lambda' \over r}-{1 \over r^2} \right)+{1 \over r^2} \label{eq12}\\
8\pi p_{reff} &=& e^{-\lambda} \left({\nu' \over r}+{1 \over r^2} \right)-{1 \over r^2} \label{eq13}\\
8\pi p_{teff} &=& {e^{-\lambda} \over 4}\left(2\nu''+\nu'^2+{2(\nu'-\lambda') \over r}-\nu' \lambda' \right) \label{eq14}
\end{eqnarray}
where,
\begin{eqnarray}
\rho_{eff} &=& \rho+{\chi \over 24\pi} \Big(9\rho-p_r-2p_t\Big) \nonumber \\
p_{reff} &=& p_r-{\chi \over 24\pi} \Big(3\rho-7p_r-2p_t \Big) \nonumber \\
p_{teff} &=& p_t-{\chi \over 24\pi} \Big(3\rho-p_r-8p_t \Big). \nonumber
\end{eqnarray}

Now the decoupled field equations (\ref{eq12})-(\ref{eq14}) are as follows:
\begin{eqnarray}
\rho &=& \frac{e^{-\lambda}}{48 r^2 (\chi +2 \pi ) (\chi +4 \pi )} \Big[r \lambda ' \left\{16 (\chi +3 \pi )-r \chi  \nu '\right\}+16 (\chi +3 \pi ) (e^{\lambda}-1) \nonumber \\
&& +r \chi  \left\{2 r \nu ''+\nu ' \left(r \nu '+4\right)\right\} \Big] \label{eq15}\\
p_r &=& \frac{e^{-\lambda}}{48 r^2 (\chi +2 \pi ) (\chi +4 \pi )} \Big[r \big\{\chi  \lambda ' (r \nu '+8)-2 r \chi  \nu ''+ \nu ' \left(20 \chi-r \chi  \nu '+48 \pi \right)\big\} \nonumber \\
&& -16 (\chi +3 \pi )(e^{\lambda}-1) \Big] \label{eq16}\\
p_t &=& \frac{e^{-\lambda}}{48 r^2 (\chi +2 \pi ) (\chi +4 \pi )} \Big[r \Big\{-\lambda ' \big\{r (5 \chi +12 \pi ) \nu '+4 (\chi +6 \pi )\big\}+2 r (5 \chi +12 \pi ) \nu '' \nonumber \\
&& +r (5 \chi +12 \pi ) \nu '~^2+ 8 (\chi +3 \pi ) \nu '\Big\}+ 8 \chi  \left(e^{\lambda }-1\right) \Big] . \label{eq17}
\end{eqnarray}
To solve the field equations, we need to assume some of the physical quantities which satisfy a strict physical constraints.

\section{Concepts of embedding class one}\label{sec4}

Kasner coordinate transformation \citep{kas21} shows that the exterior Schwarzschild vacuum is class two spacetime. Using the transformations given below
\begin{eqnarray}
X &=& {R \sin t \over \sqrt{R^2+16m^2}},~Y = {R \cos t \over \sqrt{R^2+16m^2}},~~~
Z = \int \sqrt{1+{256m^4 \over (R^2+16m^2)^3}}~dR, \nonumber
\end{eqnarray}
where $R=\sqrt{8m(r-2m)}$ and $r^2=x^2+y^2+z^2$, the well-known Schwarzschild vacuum reduces to
\begin{equation}
ds^2=-dx^2-dy^2-dz^2+dX^2+dY^2-dZ^2. \label{eq21}
\end{equation}
This means that the Schwarzschild exterior spacetime can be embedded in six dimensional pseudo-Euclidean manifold. This method was extended for general four dimensional spacetime of the form (\ref{met}) by \cite{gup75}. The chosen coordinate transformations were
\begin{eqnarray}
&& z_1=ke^{\nu/2} \cosh \left({t \over k}\right),~z_2=ke^{\nu/2} \sinh \left({t \over k}\right),~z_3=f(r), \nonumber \\
&& z^4=r \sin \theta \cos \phi ,~z_5=r \sin \theta \sin \phi,~ z_6=r \cos \theta, \nonumber
\end{eqnarray}
which transforms (\ref{met}) into
\begin{equation}
ds^2 = (dz_1)^2-(dz_2)^2 \mp (dz_3)^2-(dz_4)^2-(dz_5)^2-(dz_6)^2, \label{li1}
\end{equation}
with $[f'(r)]^2 = \mp \big[-\big(e^\lambda-1 \big)+k^2 e^\nu \nu'^2/4 \big]$.\\

Equation (\ref{li1}) also implies that the interior line element (\ref{met}) can be embedded in six dimensional pseudo-Euclidean space, however, if $(dz_3)^2=[f'(r)]^2 = 0$, it can be embedded in 5-D Euclidean space i.e.
\begin{equation}
[f'(r)]^2 = \mp \Big[-\big(e^\lambda-1 \big)+{k^2 e^\nu \nu'^2 \over 4} \Big] =0,
\end{equation}
which implies
\begin{equation}
e^\lambda = 1+{k^2 \over 4}~\nu'^2 e^\nu \label{b1}
\end{equation}
i.e. (\ref{li1}) will reduce to
\begin{equation}
ds^2 = (dz_1)^2-(dz_2)^2 -(dz_4)^2-(dz_5)^2-(dz_6)^2,
\end{equation}
a class one spacetime.

The same condition (\ref{b1}) was originally derived by \citep{kar48} in the form of components of Riemann tensor as
\begin{equation}
R_{rtrt}R_{\theta \phi \theta \phi} = R_{r\theta r \theta}R_{\phi t \phi t}+R_{r \theta \theta t}R_{r \phi \phi t}. \label{karma}
\end{equation}
\cite{pan81} pointed out that Karmarkar condition is only the necessary condition to become a class one, they discovered the sufficient condition as $R_{\theta \phi \theta \phi}\ne 0$. Hence, the necessary and sufficient condition to be a class one is to satisfy both Karmarkar and Pandey-Sharma conditions. In terms of the metric components, (\ref{karma}) can be written as
\begin{eqnarray}
{2\nu'' \over \nu'}+\nu'={\lambda' e^\lambda \over e^\lambda -1}
\end{eqnarray}
which on integration one gets the
\begin{equation}
e^{\nu}=\left(A+B\int \sqrt{e^{\lambda}-1}~dr\right)^2.\label{eq27}
\end{equation}
where $A$ and $B$ are constants of integration. In general relativity there is no class one exterior as the existing Schwarzschild's exterior itself is a class two spacetime. It has been shown that the two class one isotropic-neutral solutions in general relativity i.e. Schwarzschild uniform density model and Kohler-Chao infinite boundary model are the only two possible solutions \citep{singh17}. However, \cite{mus20} have also shown that these two isotropic-neutral solutions still exist in $f(R,T)-$theory as well. Although, the constant density model in GR can have decreasing density in $f(R,T)-$gravity and the infinite boundary Kohler-Chao solution can have finite boundary where the pressure vanishes, due to the $f(R,T)-$term.

%%%%%%%%%%%%
\begin{figure}[tbp]
\centering
\includegraphics[width=.45\textwidth,origin=c,angle=0]{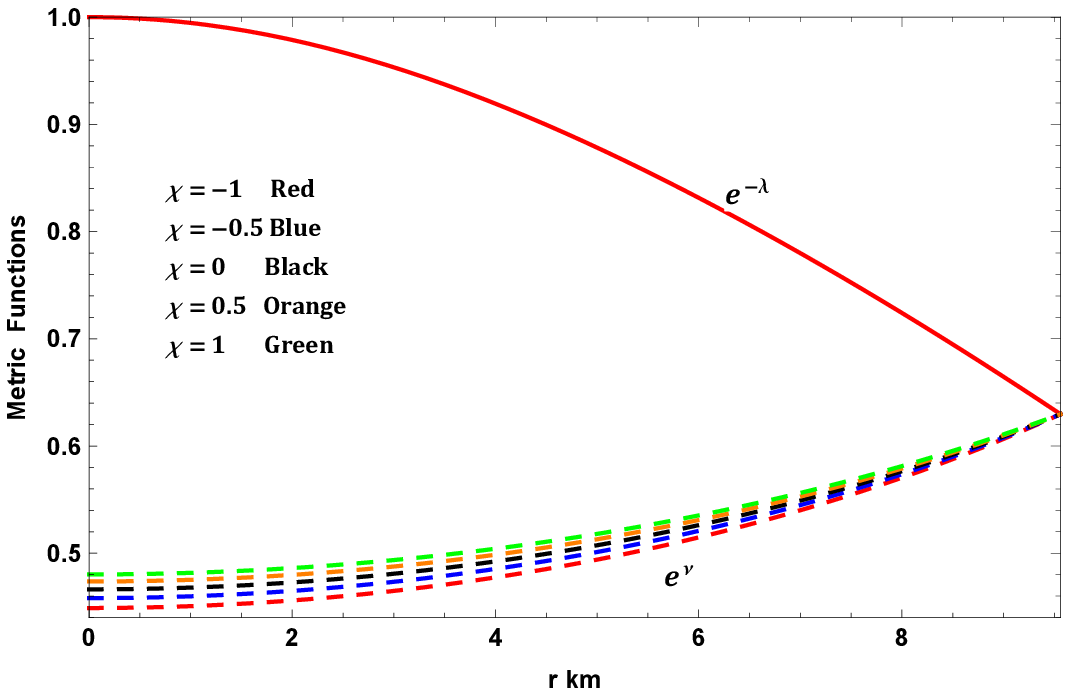}
\hfill
\includegraphics[width=.45\textwidth,origin=c,angle=0]{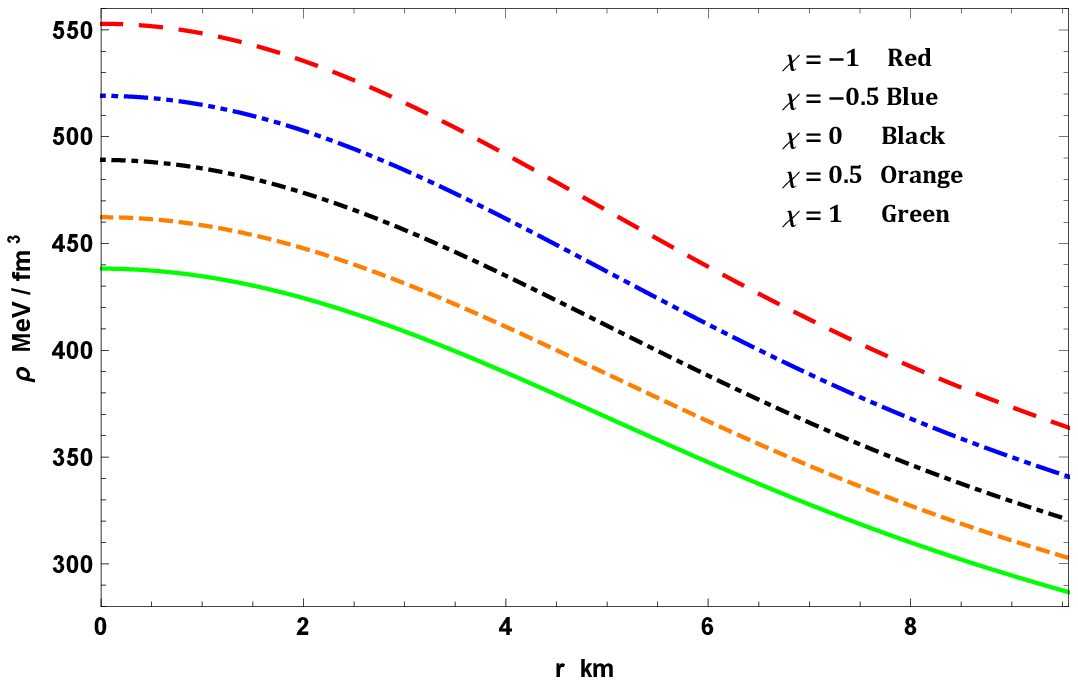}
\caption{\label{f1} Variation of metric functions and density with radial coordinate for Vela X-1 ($M = 1.77 \pm 0.08 ~M_\odot,~ R =  9.56 \pm 0.08 ~km$) with $b=0.0005/km^2$ and $c = 0.000015/km^4$.}
\end{figure}
%%%%%%%%%%%

%%%%%%%%%%%%
\begin{figure}[tbp]
\centering
\includegraphics[width=.45\textwidth,origin=c,angle=0]{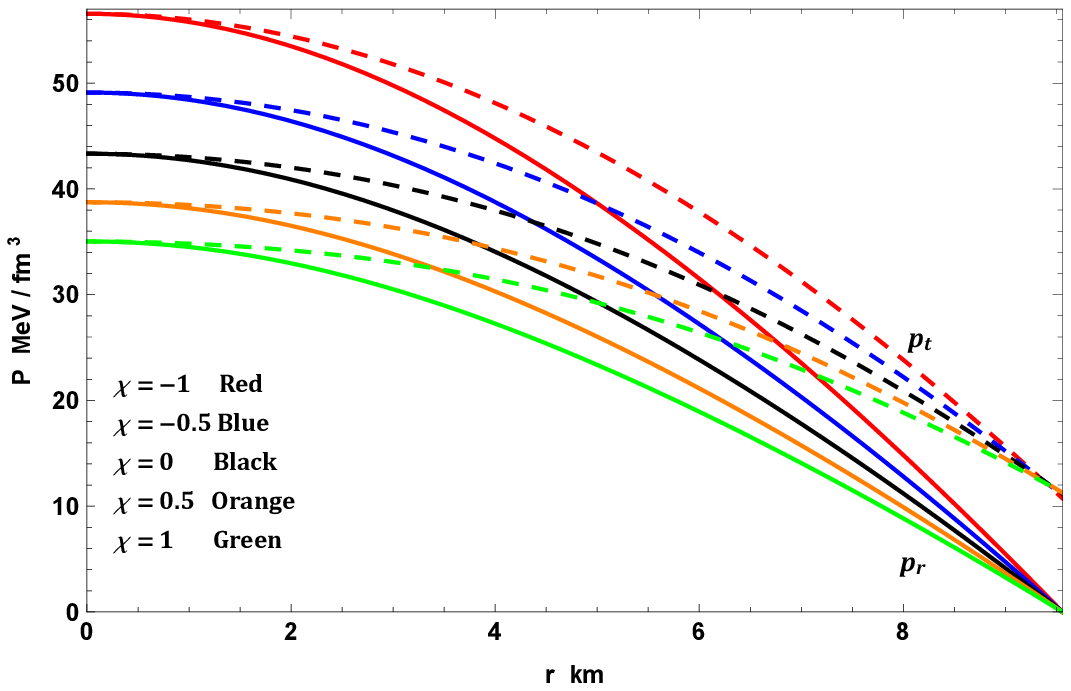}
\hfill
\includegraphics[width=.45\textwidth,origin=c,angle=0]{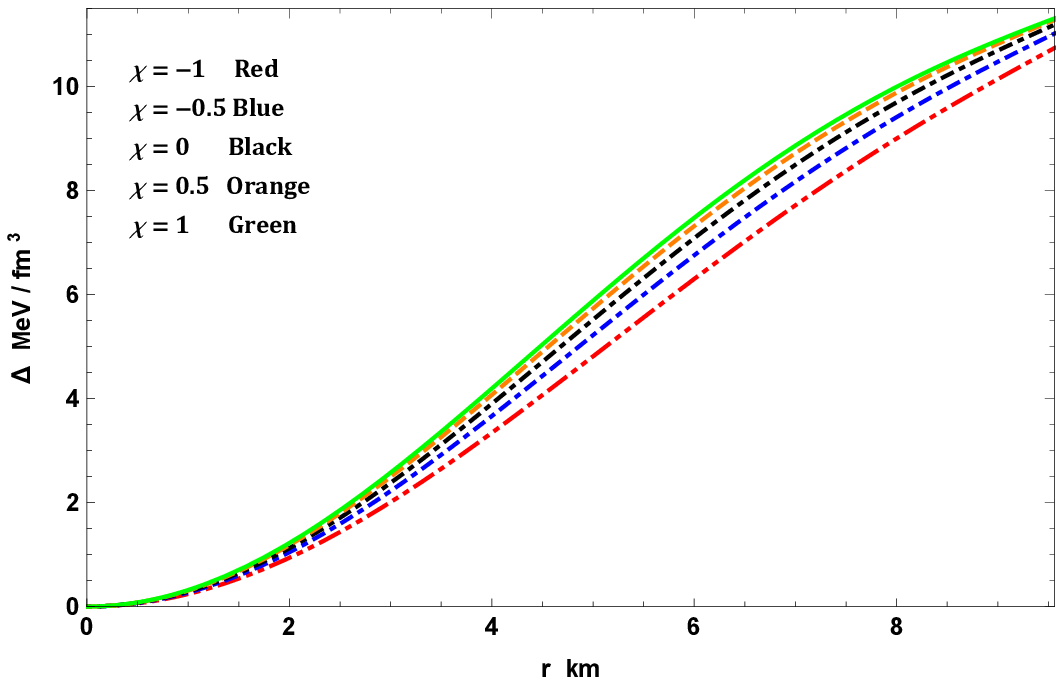}
\caption{\label{f2} Variation of pressures and anisotropy with radial coordinate for Vela X-1 ($M = 1.77 \pm 0.08 ~M_\odot,~ R =  9.56 \pm 0.08 ~km$) with $b=0.0005/km^2$ and $c = 0.000015/km^4$. }
\end{figure}
%%%%%%%%%%%

\section{Embedding class one background in $f(R,T)-$gravity}\label{sec5}

Solving the field equations in $f(R,T)-$gravity exactly is a challenging task because of the highly coupled non-linear differential equations. To simplify the problem, we have adopted the embedding class one approach, which is application to all four dimensional spacetime. Here, we propose a new metric function
\begin{equation}
e^\lambda = 1+a r^2 e^{b r^2+c r^4}. \label{eq29}
\end{equation}

When ansatz $e^\lambda$, one must keep is mind that it must be increasing function of radial coordinate and unity at the center. Using (\ref{eq29}) in (\ref{eq27}), we get
\begin{eqnarray}
e^\nu &=& \left[A+\frac{B}{\sqrt{2 c}}~\sqrt{a e^{b r^2+c r^4}}~F\left(\frac{2 c r^2+b}{2 \sqrt{2c}}\right)\right]^2\label{eq30}
\end{eqnarray}
where $F(x)$ is the Dawson's integral defined by
\begin{equation}
F(x) = e^{-x^2}\int_0^x e^{\tau^2}~d\tau = {\sqrt{\pi} \over 2} ~e^{-x^2}~\mbox{erfi}(x). \nonumber
\end{equation}

Here $\mbox{erfi}(x)$ is the usual imaginary error function. The variations of the metric functions are shown in Fig. \ref{f1} (left).\\

Plugging the metric functions into the field equations (\ref{eq15})-(\ref{eq17}), one can write
\begin{eqnarray}
\rho &=& \frac{\sqrt{a e^{b r^2+c r^4}}}{6 (\chi +2 \pi ) (\chi +4 \pi ) f_3(r) \left(a r^2 e^{b r^2+c r^4}+1\right)^2}\Big[(\chi +3 \pi ) f_2(r) \nonumber \\
&&  F\left(\frac{2 c r^2+b}{2 \sqrt{2c}}\right)+\sqrt{c} f_1(r)\Big] \label{eq31}\\
p_r &=& \frac{\sqrt{a e^{b r^2+c r^4}}}{6 (\chi +2 \pi ) (\chi +4 \pi ) f_3(r) \left(a r^2 e^{b r^2+c r^4}+1\right)^2} \bigg[\sqrt{c} f_4(r) + 2 \sqrt{2} a B f_5(r) \nonumber \\
&&  e^{b r^2+c r^4} F\left(\frac{2 c r^2+b}{2 \sqrt{2c}}\right)\bigg]\label{eq32}
\end{eqnarray}
\begin{eqnarray}
\Delta &=& \frac{r^2 \sqrt{a e^{b r^2+c r^4}} \left(a e^{b r^2+c r^4}-b-2 c r^2\right)}{2 (\chi +4 \pi ) f_3(r) \left(a r^2 e^{b r^2+c r^4}+1\right)^2} \bigg[F\left(\frac{2 c r^2+b}{2 \sqrt{2c}}\right) \sqrt{2} a B e^{b r^2+c r^4}  \nonumber \\
&&  + 2 \sqrt{c} \left(A \sqrt{a e^{b r^2+c r^4}}-B\right) \bigg],\\
p_t &=& p_r+\Delta.
\end{eqnarray}

where,
\begin{eqnarray}
f_1(r) &=& 4 A (\chi +3 \pi ) \left(a r^2 e^{b r^2+c r^4}+2 b r^2+4 c r^4+3\right)\sqrt{a e^{b r^2+c r^4}} \nonumber \\
&& +B \chi  \left(2 a r^2 e^{b r^2+c r^4}+b r^2+2 c r^4+3\right) \nonumber \\
f_2(r) &=& 2 \sqrt{2} a B e^{b r^2+c r^4} \left(a r^2 e^{b r^2+c r^4}+2 b r^2+4 c r^4+3\right)  \nonumber \\
f_3(r) &=& \sqrt{2} B \sqrt{a e^{b r^2+c r^4}} F\left(\frac{2 c r^2+b}{2 \sqrt{2c}}\right)+2 A \sqrt{c} \nonumber \\
f_4(r) &=& B \bigg[\chi  \left(10 a r^2 e^{b r^2+c r^4}-b r^2-2 c r^4+9\right)+24 \pi \left(a r^2 e^{b r^2+c r^4}+1\right)\bigg] \nonumber \\
&& -4 A \sqrt{a e^{b r^2+c r^4}} \bigg[3 \pi  \bigg(1+  a r^2 e^{b r^2+c r^4}\bigg)-r^2 \chi  \left(b-a e^{b r^2+c r^4}+2 c r^2\right)\bigg] \nonumber \\
f_5(r) &=& r^2 \chi  \left(b-a e^{b r^2+c r^4}+2 c r^2\right)-3 \pi  \bigg(a r^2 e^{b r^2+c r^4} +1\bigg) . \nonumber 
\end{eqnarray}

The variations of the density, pressures, anisotropy and EoS parameter are shown in Figs. \ref{f1} (right), \ref{f2}, \ref{f3} (left).

\section{Physical analysis on the new solution}\label{sec6}
Any new solutions must be analyze through various physical tests. After satisfying all the physical constraints one can proceed further for modeling physical systems.

\subsection{Hydrostatc equilibrium}
All the physical compact stars are believed to be in equilibrium state. Such equilibrium state can be tested by using equation of hydrostatic equilibrium or the modified TOV-equation which is given by
\begin{eqnarray}
-{\nu' \over 2}(\rho+p_r)&-&{dp_r \over dr}+{2\Delta \over r}+ {\chi \over 3(8\pi+2\chi)} {d \over dr} \big(3\rho-p_r-2p_t \big) = 0.
\end{eqnarray}

\begin{figure}[tbp]
\centering
\includegraphics[width=.45\textwidth,origin=c,angle=0]{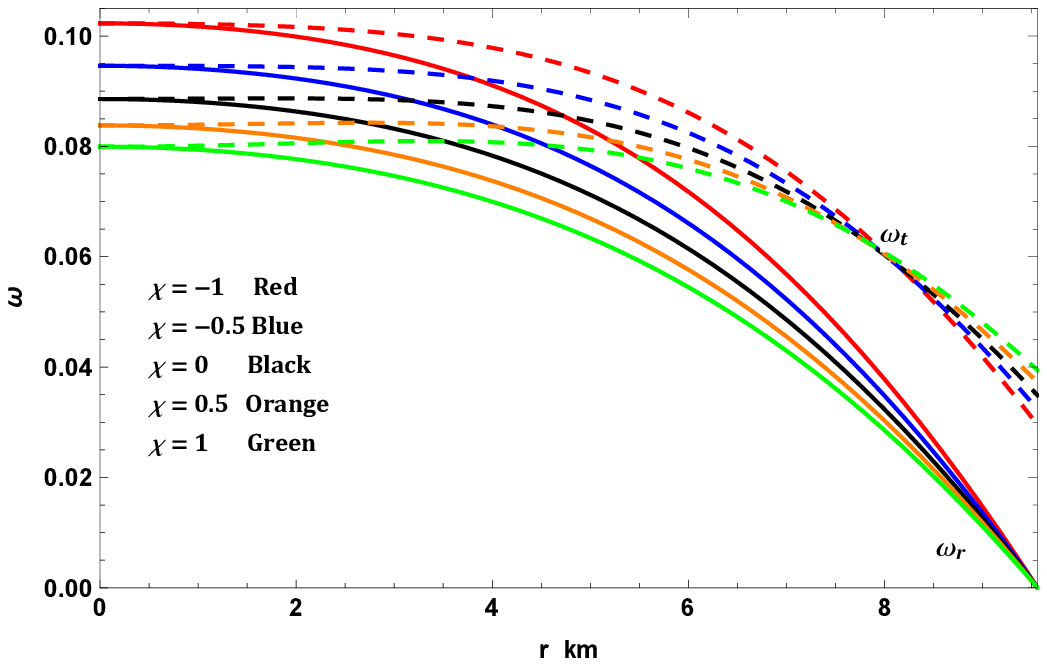}
\hfill
\includegraphics[width=.45\textwidth,origin=c,angle=0]{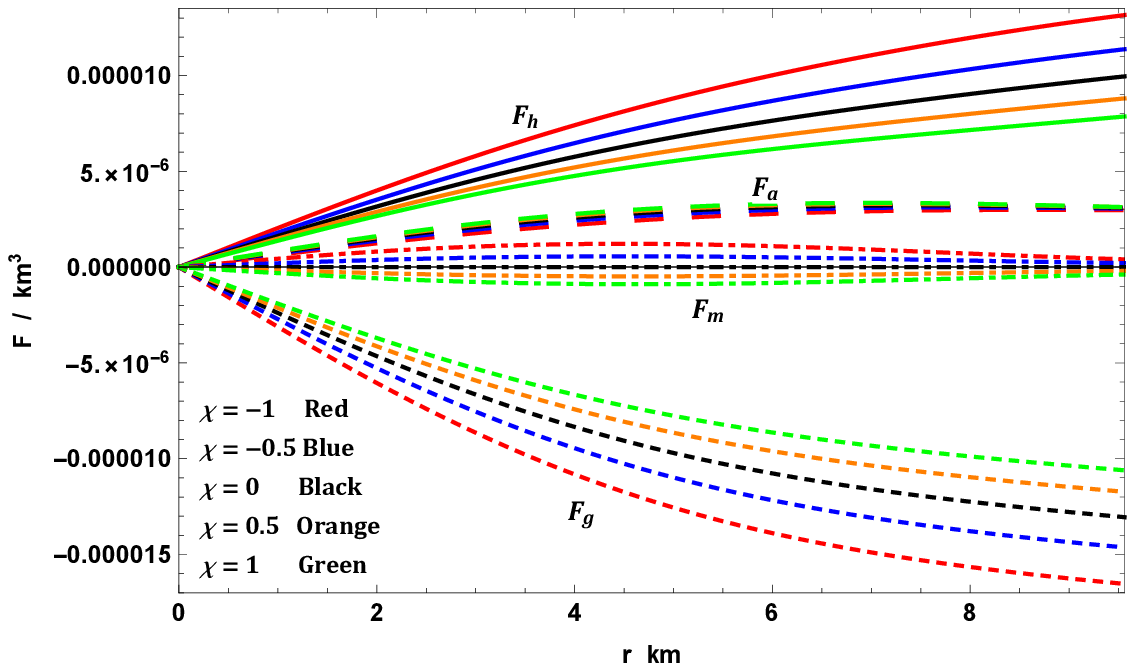}
\caption{\label{f3} Variation of equation of state parameters and forces in TOV-equation with radial coordinate for Vela X-1 ($M = 1.77 \pm 0.08 ~M_\odot,~ R =  9.56 \pm 0.08 ~km$) with $b=0.0005/km^2$ and $c = 0.000015/km^4$. }
\end{figure}

Here, the first term is gravity ($F_g$), second term is pressure gradient ($F_h$), third term is the anisotropic force ($F_a$) and the last term is the additional force ($F_m$) in $f(R,T)-$gravity. The fulfillment of the modified TOV-equation is shown in Fig. \ref{f3} (right). It shows that the forces due to gravity, pressure gradient and $F_m$ are highest in $\chi=-1$, however, anisotropic force is lowest. This will enable to hold more mass than other for lesser values of $\chi$. As $\chi$ increases the $F_g,~F_m$ and $F_h$ decreases although the $F_a$ slightly increase thereby the maximum mass that the can be hold by the system will also reduces.

%%%%%%%%%%%%
%%%%%%%%%%%

\subsection{Causality condition and stability factor}

\begin{figure}[tbp]
\centering
\includegraphics[width=.45\textwidth,origin=c,angle=0]{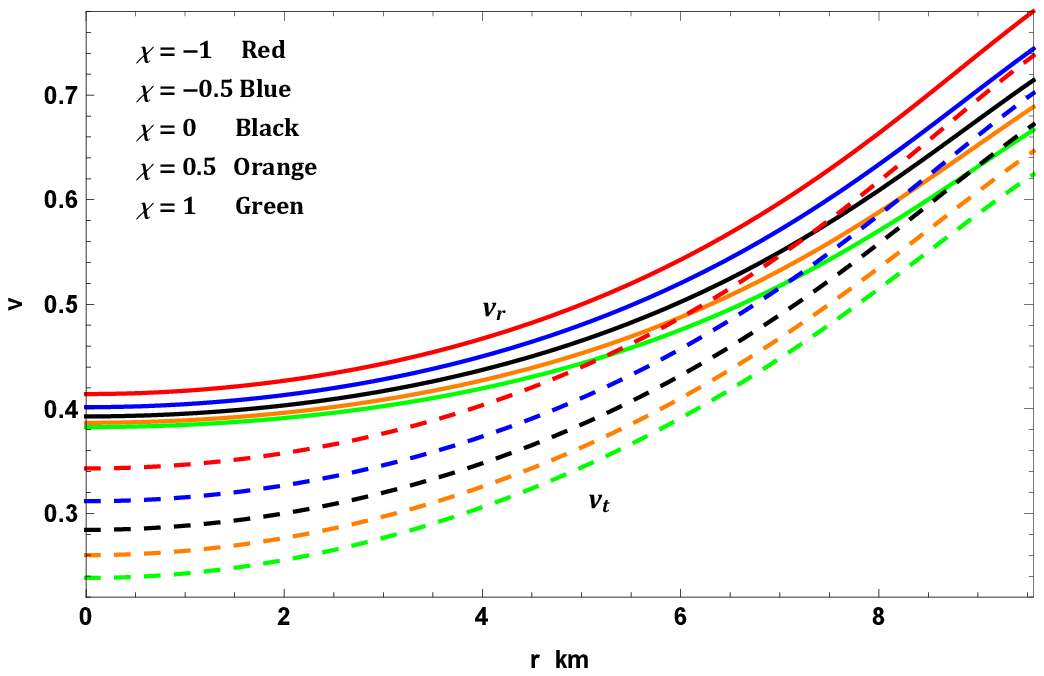}
\hfill
\includegraphics[width=.45\textwidth,origin=c,angle=0]{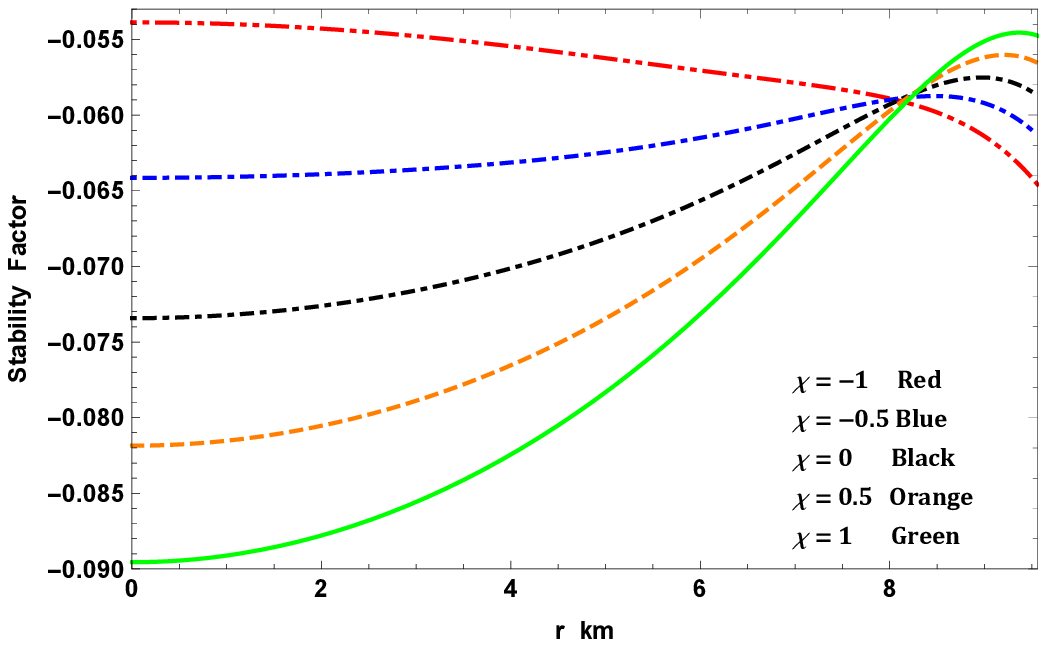}
\caption{\label{f4} Variation of sound speeds and stability factor with radial coordinate for Vela X-1 ($M = 1.77 \pm 0.08 ~M_\odot,~ R =  9.56 \pm 0.08 ~km$) with $b=0.0005/km^2$ and $c = 0.000015/km^4$. }
\end{figure}

We all aware of $f(R,T)-$gravity as an extension of general relativity, which provides a constraint on maximum speed limit. All the particle with non-zero rest mass much travel at subluminal speeds i.e. less than the speed of light (causality condition). The velocity of sound in a medium must also satisfy the causality condition and it determines the stiffness of the related EoS. Therefore, one can determine the sound speed in stellar medium to relate its stiffness. The most stiff EoS is the Zeldovich's fluid ($p_z = \rho_z$) where the sound travels exactly at light speed. The sound speed can be determine as
\begin{equation}
v_r^2 = {dp_r \over d\rho}~~,~~v_t^2 = {dp_t \over d\rho}.
\end{equation}
In Fig. \ref{f4} (left), we plot the speed of sound with the radial coordinate. It can be seen that the speed of sound is maximum for $\chi=-1$ and decreases with increase in $\chi$. This imply that the solution leads to an stiffer EoS with lesser values of $\chi$.\\

The speed of sound can also related to the stability of the configuration. As per \cite{abreu}, the stability factor can be defined as $v_t^2-v_r^2$. So long as $v_r>v_t$, the system is generally considered stable, or in other form $-1\le v_t^2 - v_r^2 \le 0$, otherwise unstable. The variation of stability factor is also shown in Fig. \ref{f4} (right) which clearly indicates the solution is stable.

%%%%%%%%%%%%
%%%%%%%%%%%

\subsection{Adiabatic index and stability}

Another parameter that determines the stability and stiffness of EoS is the adiabatic index which is defined as the ratio of specific heat at constant pressure to the specific heat at constant volume. For any fluid distribution the adiabatic index can be determine as \citep{bon64}
\begin{equation}
\gamma = {p_r+\rho \over p_r} v_r^2.
\end{equation}
As per Bondi's perceptions, the stellar fluid distribution is stable if $\gamma>4/3$ in Newtonian limit. If $\gamma \le 1$ contraction is possible and catastrophic if $\gamma < 1$. This is no longer valid for anisotropic fluids. This was extended by \cite{chann} to anisotropic fluid. For  anisotropic fluids, the stable limit of $\gamma$ depends in the nature of anisotropy and its initial configuration. If anisotropy $\Delta >0$, the stable limit will be still $\gamma>4/3$, however, if $\Delta < 0$ stability is still possible even if $\gamma < 4/3$. The variation of adiabatic index is shown in Fig. \ref{f5} (left). For different values of $\chi$ the central adiabatic index is accumulated around 2.
 
\subsection{Static stability criterion}

\begin{figure}[tbp]
\centering
\includegraphics[width=.45\textwidth,origin=c,angle=0]{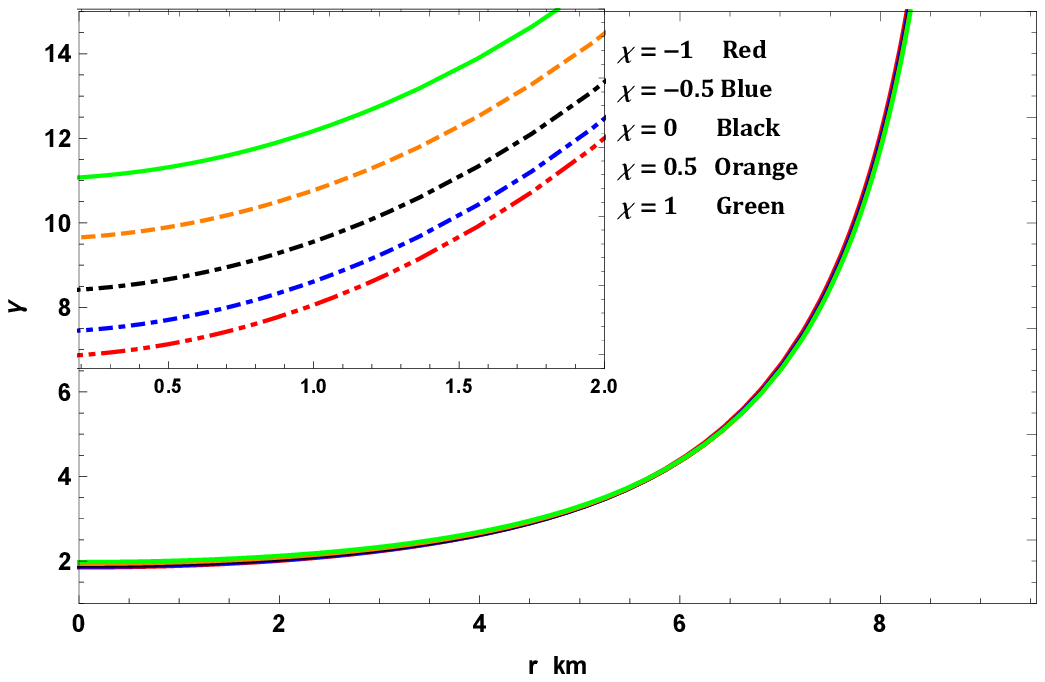}
\hfill
\includegraphics[width=.45\textwidth,origin=c,angle=0]{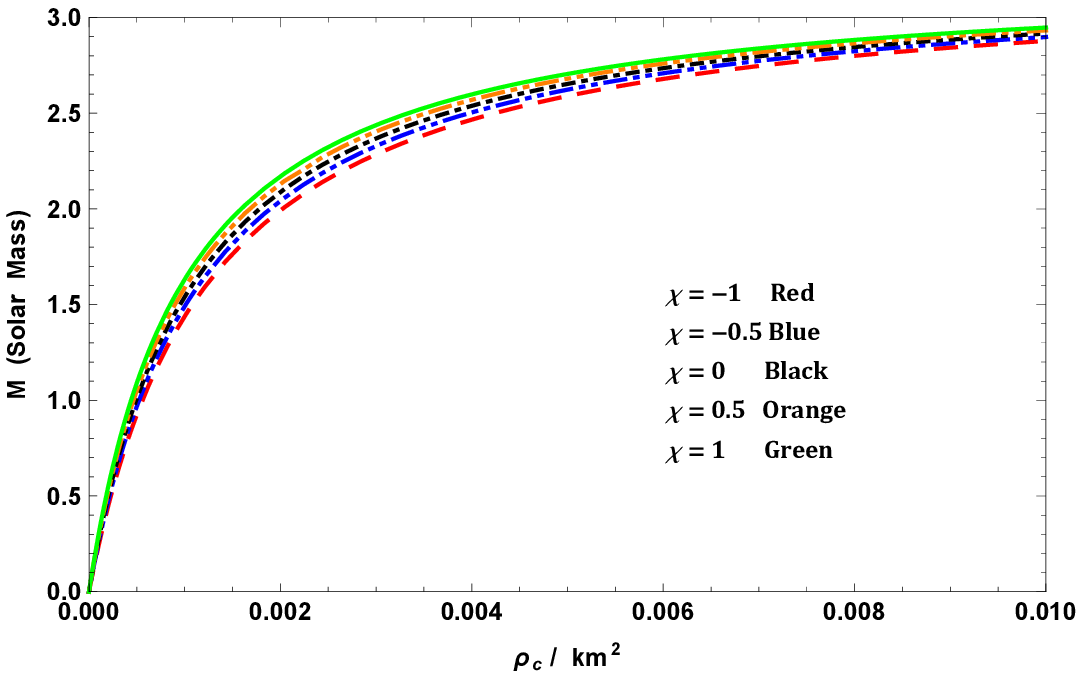}
\caption{\label{f5} Variation of adiabatic index with radial coordinate and variation of mass with central density for Vela X-1 ($M = 1.77 \pm 0.08 ~M_\odot,~ R =  9.56 \pm 0.08 ~km$) with $b=0.0005/km^2$ and $c = 0.000015/km^4$. }
\end{figure}

This criterion analyze the stability of stellar configurations under radial perturbations originally established by \cite{chand64}. Further, \cite{harri} and \cite{zeld} simplifies this method. The static stability criterion imposed the condition that if $\partial M / \partial \rho_c$ is greater than zero, the system is stable otherwise unstable. To see it, we have calculate the mass as a function of $\rho_c$ given as
\begin{eqnarray}
M\left(\rho _c\right)= {R \over 2}\left(1-{1 \over 1+a R^2 e^{b R^2+c R^4}}\right).
\end{eqnarray}
Here $a$ is a very complicated function of $\rho_c$ and therefore we avoid to mention. The variation of mass with respect to the central density is shown in Fig. \ref{f5} (right). From this, one can conclude that the stability is enhance with increase in $\chi$. This is because the range central density is more for saturating the mass when $\chi=1$ than $\chi=-1$. This implies that the stable range of density during radial oscillation is more for higher values of $\chi$. This can conclude that that solution is stable under radial perturbations.

%%%%%%%%%%%%
%%%%%%%%%%%

\subsection{Energy conditions}

\begin{figure}[tbp]
\centering
\includegraphics[width=.45\textwidth,origin=c,angle=0]{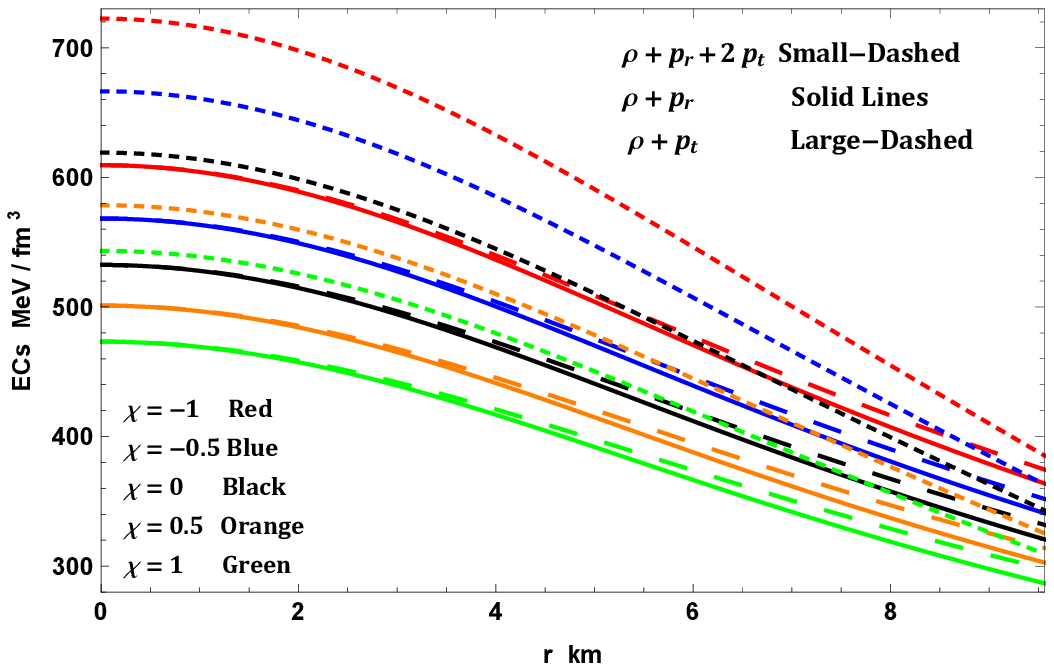}
\hfill
\includegraphics[width=.45\textwidth,origin=c,angle=0]{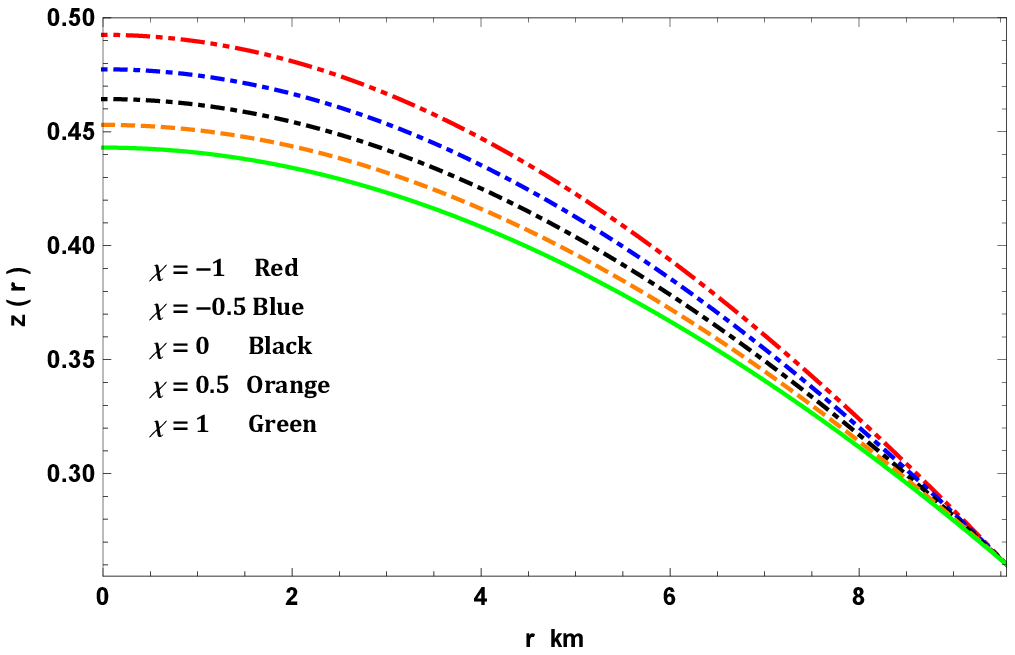}
\caption{\label{f6} Variation of energy conditions and redshift with radial coordinate for Vela X-1 ($M = 1.77 \pm 0.08 ~M_\odot,~ R =  9.56 \pm 0.08 ~km$) with $b=0.0005/km^2$ and $c = 0.000015/km^4$. }
\end{figure}

After confirming all the stability tests, the nature of  matter content i.e. either normal (baryonic, hadronic etc.) or exotic (dark matter, dark energy etc.) can be identified by using energy conditions. Satisfaction or violation of certain energy conditions will imply the nature of the matter. These energy conditions are given as
\begin{eqnarray}
\mbox{Null} &:& \rho+p_r \ge 0 ,~\rho+p_t \ge 0, \nonumber \\
\mbox{Weak} &:& \rho+p_r \ge 0 ,~\rho+p_t \ge 0, ~\rho \ge 0, \nonumber \\
\mbox{Strong} &:& \rho+p_r \ge 0 ,~\rho+p_t \ge 0, ~\rho+p_r+2p_t \ge 0, \nonumber \\ 
\mbox{Dominant} &:& \rho \ge |p_r| ,~\rho \ge |p_t|. \nonumber
\end{eqnarray}\\
From Fig. \ref{f6} (left), it is found that all the energy conditions are satisfied by the solution and therefore, the matter content is normal.

\section{Boundary conditions}\label{sec7}

Boundary conditions ensure that the interior and exterior spacetime are connected. As usual, we will assume the exterior as Schwarzschild vacuum given as
\begin{eqnarray}
ds_+^2 &=& \left(1-{2m \over r}\right) dt^2 - \left(1-{2m \over r}\right)^{-1}dr^2 - r^2(d\theta^2 + \sin^2 \theta ~d\phi^2).
\end{eqnarray}
However, we must keep in mind that to avoid singularity, one must satisfy $r>2m$.

At the surface $r=R$, we get $ds_-^2|_{r=R}=ds_+^2|_{r=R}$ which imply
\begin{eqnarray}
e^{-\lambda(R)} = 1-{2M \over R} = e^{\nu(R)}. \label{eq42}
\end{eqnarray}

%%%%%%%%%%%%
%%%%%%%%%%%

On using (\ref{eq42}), we get
\begin{eqnarray}
a &=& \frac{2 M e^{-R^2 \left(b+c R^2\right)}}{R^2 (R-2 M)} \\
A &=& \sqrt{1-\frac{2 M}{R}}-\frac{B}{\sqrt{2 c}} \sqrt{a e^{b R^2+c R^4}}~ F\left(\frac{2 c R^2+b}{2 \sqrt{2 c}}\right).
\end{eqnarray}
Generally, in modeling compact stars the pressure at the surface needs to vanish i.e. $p_r(R)=0$. This condition allow us to determine one more constant as
\begin{eqnarray}
B &=& 4 \sqrt{c} \sqrt{1-\frac{2 M}{R}} \sqrt{a R^2 e^{b R^2+c R^4}} \bigg[R^2 \chi  \Big(2 c R^2-a e^{b R^2+c R^4}+b\Big)-3 \pi  \left(a R^2 e^{b R^2+c R^4}+1\right)\bigg] \nonumber \\
&&  \bigg[\sqrt{c} R \Big\{\chi  \Big(b R^2+2 c R^4-9  -10 a R^2 e^{b R^2 +c R^4}\Big)-24 \pi  \Big(a R^2 e^{b R^2+c R^4}+1\Big)\Big\}+2 \sqrt{2} a R \nonumber \\
&& e^{b R^2+c R^4} F\left(\frac{2 c R^2+b}{2 \sqrt{2c}}\right) \Big\{3 \pi  \left(a R^2 e^{b R^2+c R^4}+1\right)-R^2 \chi \left(b-a e^{b R^2+c R^4}+2 c R^2\right)\Big\} \nonumber\\
&& -{2aR \sqrt{2} \over e^{-(b R^2+c R^4)}} F\left(\frac{2 c R^2+b}{2 \sqrt{2c}}\right)  \bigg\{3 \pi  \left(a R^2 e^{-(b R^2+c R^4)}+1\right) -R^2 \chi \nonumber\\
&&  \hspace{2 cm} \left(b-a e^{-(b R^2+c R^4)}+2 c R^2\right)\bigg\}\bigg]^{-1}.
\end{eqnarray} 
The parameter $b$ and $c$ will be treated as free whereas $M$ and $R$ will be taken from the observational evidences.

\section{Stiffness of EoS, $M-R$ and $I-M$ curve}\label{sec8}

\begin{figure}[tbp]
\centering
\includegraphics[width=.45\textwidth,origin=c,angle=0]{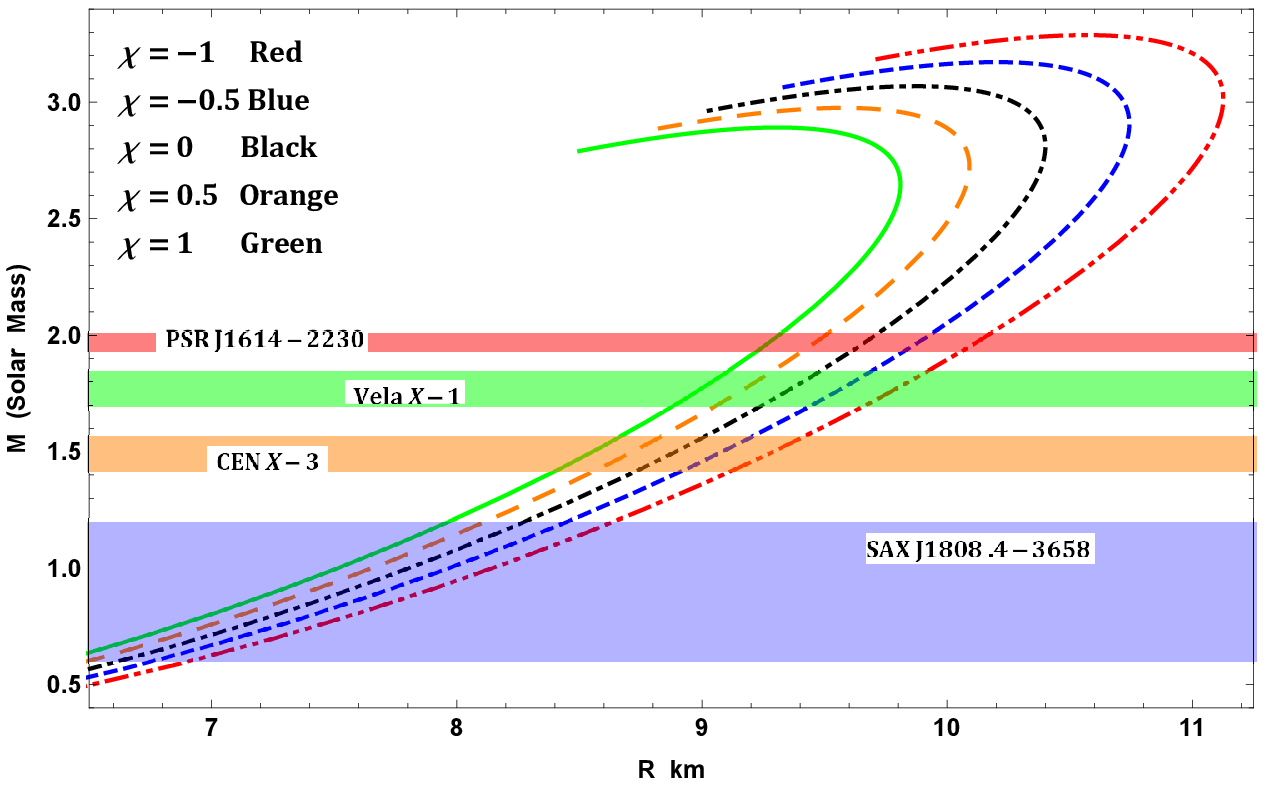}
\hfill
\includegraphics[width=.45\textwidth,origin=c,angle=0]{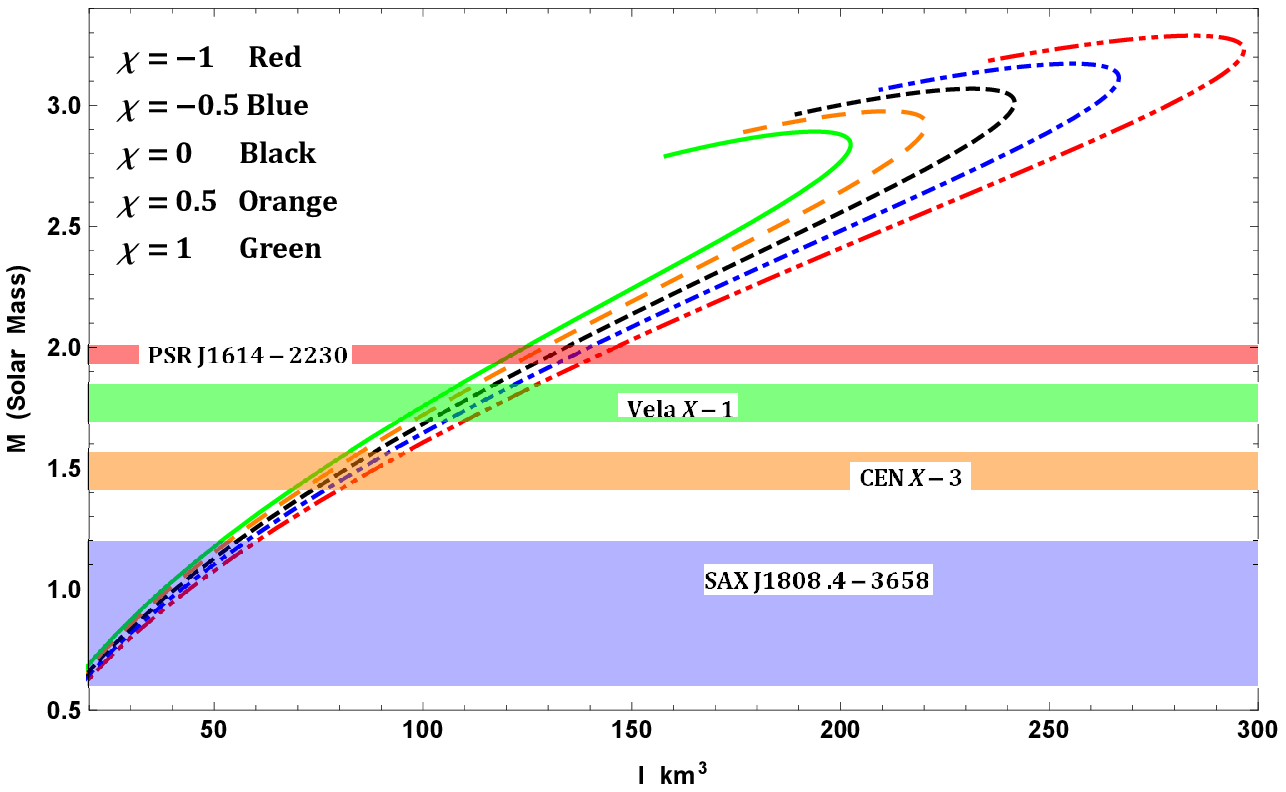}
\caption{\label{f7} Variation of total mass with radius and total with moment of inertia.}
\end{figure}

There are several ways of determining the stiffness of an EoS e.g. by determining adiabatic index, sound speed etc. However, the sensitivity to stiffness is found to be very sharp in $M-R$ and $M-I$ graphs. In fact, $M-I$ graph is the most effective and sensitive to the stiffness of an EoS. In Fig. \ref{f7} (left) we shown the variation of mass with respect to the radius. Since, from the above sections we have already noted that the EoS is stiffest for $\chi=-1$ and as $\chi$ increases the stiffness reduces. Due to this, the mass that can hold by the corresponding EoS will also be reduces as $\chi$ increases. The same nature can be seen from the $M-R$ curve in Fig. \ref{f7} (left). To compare with the $M-I$ curve, one must establish how to determine the moment of inertia ($I$). Adopting the \cite{bejg} formula one can determine the $I$ corresponding to a static solution. It is given by 
\begin{equation}
I = {2 \over 5} \left(1+{(M/R)\cdot km \over M_\odot} \right)~MR^2
. \label{eq46}
\end{equation}
The change in $I$ with respective to mass is shown in Fig. \ref{f7} (right). Again, we can verify that the EoS is most stiff for lesser values of $\chi$. The transition at the peak in $M-I$ curve is sharper than in $M-R$ curve indicates the sensitivity to the stiffness of the EoS.

Further, our generated $M-R$ curve is also fit with observational results for few well-known compact stars. As examples, we have matched for PSR J1614-2230, Vela X-1, Cen X-3 and SAX J1808.4-3658. Since the $M-R$ curve fit with these compact stars, one possibility arises from $M-I$ curve to predict the possible range of $I$ for the above mentioned objects.

%%%%%%%%%%%%
%%%%%%%%%%%

\section{Discussion and conclusion}\label{sec9}

In this article, we have successfully devised the method of embedding class one into $f(R,T)-$ gravity. This procedure not only simplify in exploring new exact solutions within $f(R,T)-$ theory but also to investigate the theory of compact stars in the same realm. The solution was analyzed through various physically stringent conditions like causality condition, energy conditions, satisfaction of TOV-equation, staiblity criteria via Bondi's condition, Abreu et al. condition, static stability criterion etc. As we increase the coupling constant $\chi$ from $-1$ to 1, the density, pressures, $\omega$, speed of sound and interior redshift increases. This increase in energy density may leads to generation of exotic particles such as quarks, hyperons \citep{zdu04,shm99}, kaon condensation \citep{lim14} that soften the equation of state. Therefore, in $M-R$ and $M-I$ curves one can see that $M_{max}$ and $I_{max}$ increases with decrease in coupling constant, which is the direct consequence of stiffening the equation of state at low $\chi$. This outcomes can also be cross check with the velocity of sound at the interior. From Fig. \ref{f4} (left), we can see that $v_{r/t}(\chi=-1)$ is always greater than $v_{r/t}(\chi=1)$, i.e. the equation of state is stiffer for the former than the later. 

Although the stiffness of the EoS is enhanced by the low coupling constant, however, the stability is compromised. In Fig. \ref{f5} (left) the central values $\Gamma_c(\chi=1)=1.977$ is more than at $\Gamma_c(\chi=-1)=1.848$. For $\chi=-1$, the value $\Gamma_c(\chi=-1)=1.848$ is comparatively closer to the Bondi limit i.e. $\Gamma=1.333$. Therefore, $\chi=-1$ is more sensitive towards radial oscillations and hence the stability. On the other hand, the range of central density is less for $\chi=-1$ than $\chi=1$. This means that, during radial perturbations the range of stable density perturbation is more for $\chi=1$ than $\chi=-1$ thereby enhancing its stability. Further, the solution fulfilled all the energy conditions. The anisotropy decreases with decrease in coupling parameter. This leads to the conclusion that anisotropy in pressure reduces with increase in stiffness of the EoS. The surface redshifts predicted form the solution in far within the Ivanov's limit i.e. $z_s=3.842$ \citep{iva02}. 

In Table \ref{tab}, we have presented 4 compact stars and few their corresponding physical parameters. Here we have also provide how the radius, central and surface densities, central pressure and moment of inertia varies with $f(R,T)-$coupling constant. Since the stiffness increases with decrease in $\chi$, the central and surface densities, central pressure and radius increases while the stability is compromised. For $-1\le \chi \le 1$, we have predicted the radii of the compact stars. All the above good results and in agreement with the observational values of masses and radii, one can undoubtedly conclude that the solution might have astrophysical significance.

For the graphical test and investigation of the physical reasonable grounds of the accomplished solutions, we have well-respected the physical profile of four well-known  compact stellar systems viz., PSR J1614-2230, Vela X-1, Cen X-3 and SAX J1808.4-3658. In this regard, we have supposed the radius-radius element of the metric function ($e^\lambda$) in the new form $e^\lambda = 1+a r^2 e^{b r^2+c r^4}$ and provided the time-time element of the metric function ($e^\nu$) as expressed explicitly in Eq.~(\ref{eq30}), as well as exhibiting the anisotropic impacts on the physical systems imposed by the $\chi-$coupling constant of the $f(R, T)$ gravity theory. As establish the most important salient features that describe the stellar system fulfills all the general necessities to guarantee a respectful framework.

The comportment of physical amounts of time-time and radius-radius components, namely, $e^\nu$ and $e^\lambda$, respectively, regarding the radial coordinate $r$, for Vela X-1 ($M = 1.77 \pm 0.08 ~M_\odot,~ R =  9.56 \pm 0.08 ~km$) are represented in Fig. \ref{f1} (left). This figure shows that both metric functions are limited at the origin and monotonically expanding towards the point of confinement at the surface. Moreover, it very well may be seen from the Eqs.(\ref{eq29}) and (\ref{eq30}) that $e^\lambda (r=0)= 1$ and $e^\nu (r=0)\neq 0$ which shows that this stellar system is realistic and agreeable. Hence, Figs. \ref{f1} (right) and \ref{f3} clearly show that all the thermodynamic observable, i.e energy density $\rho$, radial pressure $p_r$ and transverse pressure $p_t$ are well-defined within the stellar structure. In this respect, is worth mentioning that all quantities mentioned have their maximum values at the core of the stellar structure and monotonically decreasing comportment with increasing radius towards the surface. At this stage it merits referencing that the present stellar model shows a positive anisotropy factor $\Delta$, it can be observed in Fig. \ref{f2} (right) where $p_t>p_r$ then $\Delta >0$. In this way the stellar structure is uncovered to a repulsive force that neutralizes the gravitational slant, this reality permits the building of a progressively compact stellar configuration. We affirm again from Figs. \ref{f1} and \ref{f2}, that the stellar system is absolutely without physical or geometrical singularities for all chosen values of $\chi -$coupling constant of the $f(R,T)-$gravity theory. Fig. \ref{f2} (right) exhibit the variation of anisotropy parameter $\Delta$ with radial coordinate for Vela X-1. It vanishes at the center, and consequently, positive defined and increasing function towards the stellar structure surface. Additionally Fig. \ref{f3} (left) manifest that the EoS parameters viz., $\omega_r = p_r /\rho$ and $\omega_t = p_t /\rho$ are under 1, demonstrating Zeldovich’s condition is fulfilled wherever within the stellar structure in the context of $f(R, T)-$gravity theory. 

Besides, equilibrium study of the model for stellar system is established using generalized TOV-equation originates from the modified type of the energy conservation equation for the energy-momentum tensor in the arena of the $f(R,T)-$gravity theory given in Eq. (\ref{eq11}). In this regard, it is easy to see from Fig. \ref{f3} (right) that the modified TOV-equation permits the exploration under various forces that perform on the stellar structure, in this event, the stellar structure is under four various forces, namely, gravity ($F_g$), pressure gradient ($F_h$), the anisotropic force ($F_a$) and the additional force ($F_m$) in $f(R,T)-$gravity. The forces due to gravity, pressure gradient and the additional term are highest in $\chi = -1$, however, anisotropic force is lowest. This will enable to hold more mass than other for lesser values of $\chi$. As $\chi$ increases the $F_g$, $F_m$ and $F_h$ decreases although the $F_a$ slightly increase thereby the maximum mass that the can be hold by the system will also reduces. On the other hand, we investigate the stability of realistic and compact stellar structure solutions via causality condition and stability factor, stability criteria via Bondi’s condition and Harrison-Zeldovich-Novikov static stability criterion corresponding to $\chi-$coupling constant of the $f(R, T)-$gravity. According to these criteria, In Fig. \ref{f4} (left) the performances of the radial ($v_r$) and transverse ($v_t$) speed of sound with respect to the radial coordinate $r$, for the compact stellar configuration have been described and it is remarked clearly that they remain within their predetermined range $]0,1[$ throughout the stellar framework, which affirms the causality condition and furthermore valid the acceptability of the subsequent anisotropic solution of our stellar system. Moreover, It can be seen that the speed of sound is maximum for $\chi =-1$ and decreases with increase in $\chi$, which implies that our solution leads to an stiffer EoS with lesser values of $\chi -$coupling constant. The solution can also appear for static and stable astrophysical structures as the stability factor can be defined as $v_t^{2}-v_r^{2}$ lies between the bounds $-1$ to $0$ for different values of $\chi -$coupling parameter, which are shown in Fig. \ref{f4} (right). Furthermore, for a non-collapsing stellar fluid distribution, the adiabatic index should also be greater than $4/3$ for $\Delta>0$ according to stability criteria via Bondi’s perceptions, which can be observed clearly from Fig. \ref{f5} (left), so our stellar system is generally stable. In Fig. \ref{f5} (right),  we plot the variation of mass with respect to the central density which satisfies the Harrison-Zeldovich-Novikov static stability criterion. From this figure, one can conclude that the stability is enhance with increase in $\chi$. This is on the grounds that the range central density is more for saturating the mass when $\chi = 1$ than $\chi = -1$. This infers that the stable range of density during radial oscillation is more for higher values of $\chi$. This can conclude that our solution is completely stable under radial perturbations. Furthermore, we have investigated the profile of the thermodynamic quantities that prompts a well-behaved and positive defined energy-momentum tensor throughout interior the compact stellar structure which satisfies at the same time by the inequalities that governs them named ECs. Hence, In Fig. \ref{f6} (left), we have plotted the L.H.S of these inequalities which checks that all the ECs are achieved at the astrophysical inside and consequently corroborates that the physical accessibility of the compact stellar inside solution. The fulfillment of the redshift with radial coordinate $r$, for Vela X-1 is illustrated in Fig. \ref{f6} (right). From this figure, it can see that the surface redshift within typical value resulting by Ivanov’s which strongly proves the agreement of our compact stellar system. 

On the other hand, we have generated the $M-R$ curves from our solutions in the arena of $f(R, T)$ gravity theory and we found a perfect fit for a certain compact stellar spherical systems such as PSR J1614-2230, Vela X-1, Cen X-3 and SAX J1808.4-3658. Therefore, we have predicted the corresponding radii and their respective moment of inertia from the $M-I$ curve by varying the coupling constant $\chi$ as a free variable. Furthermore, the $M-R$ and $I-M$ curves are represented in Figs. \ref{f7}. From these curves, one can approve that our solution predicted the radii in good agreement with the observational data.

Finally, we wish to remark that all anisotropic spherically symmetric solutions establish in this work satisfying obtained well-behaved stellar interiors in the area of $f(R, T)$ gravity theory by using the embedding class I procedure are fulfilling and sharing all the physical and mathematical features necessary in the study of compact stellar spherical systems, which provide to understand the evolution of realistic compact stellar spherical systems. In this regard, the $f(R, T)$ gravity theory is a promising scenario to envisage the existence of compact stellar spherical systems performed by an anisotropic matter distribution, whose effects can be contrasted with the well-described GR, and meets the notable and tried general necessities.

%%%%%%%%%%%%%%%%%%%
\begin{table*}
\footnotesize
\caption{\label{tab} Prediction of radius for few well-known compact stars and their corresponding central densities and pressures for different values of $\chi$.}
\centering
\begin{tabular}{|c|c|c|c|c|c|c|c|p{0.55in}}
\hline
\hline
Objects & $\chi$ & $M$ & Predicted  & $\rho_0$ & $\rho_s$ & $p_0$ & $I \times 10^{44}$  \\
&  & ($M_\odot$) & Radius (km) & ($MeV/fm^3$) & ($MeV/fm^3$) & ($MeV/fm^3$) & ($g~cm^2$) \\
\hline
               & -1.0    &                 & 10.13  & 621.11 & 376.32 & 72.57 & 18.51\\
	           & -0.5    & 	               & 9.89   & 581.58 & 353.52 & 63.20 & 17.52\\
PSR            &  0      & 1.97            & 9.66   & 549.65 & 330.71 & 56.10 & 16.77\\
J1614-2230	   &  0.5    & $\pm$ 0.04      & 9.47   & 519.24 & 313.98 & 50.28 & 16.04\\
	           &  1.0    &                 & 9.28   & 491.87 & 295.74 & 45.44 & 15.59\\
\hline	             
           	   & -1.0	 &	                & 9.76  & 551.66 & 362.96 & 56.48 & 15.29\\
	           & -0.5    &	                & 9.55  & 518.29 & 339.95 & 49.12 & 14.62\\
VELA X-1       &  0      & 1.77             & 9.34  & 488.38 & 319.24 & 43.36 & 13.94\\
	           & 0.5     &  $\pm$ 0.08      & 9.15	& 461.91 & 303.13 &	38.76 & 13.56\\
	           &  1.0    &                  & 8.99  & 437.75 & 285.87 & 34.85 & 13.01\\
\hline	             
                 & -1.0	&             &	7.91 & 431.10 &	339.05 & 20.76 & 4.74\\
	             & -0.5	&             &	7.69 & 403.98 &	317.68 & 18.14 & 4.51\\
SAX              & 0    & 0.9         & 7.54 & 380.96 & 298.78 & 16.12 & 4.34\\
J1808.4-3658	 & 0.5	& $\pm$ 0.3    &	7.42 & 359.60 &	282.34 & 14.50 & 4.17\\
	             & 1.0	&             &	7.26 & 340.69 &	266.73 & 13.19 & 4.00\\
\hline	             
                & -1.0  &               & 9.29 & 502.10 & 352.97 & 41.72 & 11.24\\
	            & -0.5	&               & 9.03 & 471.03 & 330.02 & 36.47 & 10.85\\
CEN X-3	        & 0	    & 1.49          & 8.85 & 443.42 & 311.65 & 31.83 & 10.42\\
	            & 0.5	& $\pm$ 0.08     & 8.66 & 418.10 & 294.44 & 28.60 & 10.11\\
                & 1.0	&               & 8.53 & 397.39 & 278.37 & 25.98 & 9.72\\
\hline
\hline
\end{tabular}
\end{table*}
%%%%%%%%%%%%%%%%

%\acknowledgments

%This is the most common positions for acknowledgments. A macro is available to maintain the same layout and spelling of the heading.


\end{document}